\documentstyle[12pt]{article}

\def\be{\begin{equation}}
\def\ee{\end{equation}}
\def\bea{\begin{eqnarray}}
\def\eea{\end{eqnarray}}

\def\goto{\longrightarrow}
\def\sa{\hspace{0.1in}}

\def\sb{\hspace{0.2in}}

\def\M{{\cal M}}

\begin{document}
\oddsidemargin 5mm
\setcounter{page}{0}
\renewcommand{\thefootnote}{\fnsymbol{footnote}}
\newpage
\setcounter{page}{0}
\begin{titlepage}
\begin{flushright}
SISSA-ISAS 123/92/FM
\end{flushright}
\vspace{0.5cm}
\begin{center}
{\large {\bf A Multi-Grid Method for the Resolution of Thermodynamic
Bethe Ansatz Equations
}} \\
\vspace{1.5cm}
{Alfio Borz\`{\i} and Anni Koubek}\\
\vspace{0.8cm}
{\em SISSA, Scuola Internazionale Superiore di Studi Avanzati\\
Via Beirut 2-4, \\
34013 Trieste, \\ Italy} \\
\end{center}
\vspace{6mm}
\begin{abstract}
We present a multi-grid algorithm in order to solve numerically the
 thermodynamic Bethe ansatz equations. We specifically adapt the program to
 compute the data of the conformal field theory reached in the ultraviolet
limit.
\end{abstract}
\vspace{5mm}
{\em Program classification: 4.12-11.1}
\end{titlepage}
\newpage
{\bf \LARGE
PROGRAM SUMMARY
}
\begin{description}
\item[] {\em Title of program:} TBA
\item[] {\em Catalogue number:}
\item[] {\em Program available from:} CPC Program Library, Queen's
University of Belfast, N. Ireland
\item[] {\em Computer:} IBM RS6000/560
\item[] {\em Operating system:} AIX 3.2
\item[] {\em Programming language used:} FORTRAN 77
\item[] {\em Number of lines in program:} 1287
\item[] {\em Key words:} two-dimensional systems away from criticality,
conformal field theory, thermodynamic Bethe ansatz, multi-grid methods.
\item[] {\em Nature of the physical problem:} resolve the thermodynamic
Bethe ansatz equations corresponding to a factorized scattering theory
and extract the data of the underlying conformal field theory reached in
the ultraviolet limit.
\item[] {\em Method of solution:} non linear multi-grid method with
full adaptive scheme.
\item[] {\em Typical running time:} strongly dependent on the dimension
of the system of coupled integral equations, see Table 1.
\item[] {\em References:}
\begin{enumerate}
\item A. Brandt, {\em Multi-grid techniques: 1984 guide with
applications to fluid dynamics} (GMD-Studien. no 85,
St. Augustin, Germany, 1984).
\item W. Hackbusch,{\em Multi-grid methods and applications}
(Sprin\-ger-Ver\-lag Berlin, Heidelberg, 1985).
\item W.T. Vetterling, S.A. Teukolsky, W.H. Press, B.P. Flannery,
{\em  Numerical reci\-pes: The art of scientific computing}
(Cambridge University Press, New York, 1985).
\end{enumerate}
\end{description}

\newpage
\renewcommand{\thefootnote}{\arabic{footnote}}
\setcounter{footnote}{0}
{\bf \LARGE
LONG WRITE-UP
}
\section{Introduction}

Massive relativistic field theories can be described on-shell by their
scattering matrix. This approach is specially fruitful in two dimensions,
where there exists a large class of models which are integrable, and their
$S$-matrix can be computed exactly, being factorizable \cite{zam}.
 Unfortunately there is no general direct method in order to compute the
$S$-matrix of a theory, but usually it is conjectured from general
axioms and the underlying symmetries of the corresponding Hamiltonian.\\
The thermodynamic Bethe ansatz (TBA) was developed in order to provide a
means to link a conjectured scattering theory with the underlying field
theory \cite{zamtba}. It describes the finite temperature effects of the
factorized relativistic field theory, using the $S$ matrix as an input.
If one studies the high temperature limit of the TBA equations, one can
identify the conformal field theory (CFT) which governs the ultraviolet
behaviour of the underlying
 field theory. One should  though note, that it is not guaranteed
that every consistent $S$-matrix describes the scattering in some field
theoretical model! Therefore the axiomatic bootstrap approach is only
of limited value if not linked to field theory by some means, wherefrom
the TBA is one of the most powerful ones.\\
Given the scattering data one can in most cases extract analytically the
central charge of the CFT reached in the conformal limit, and in some
cases the dimension of the perturbing operator, if the symmetry of the
problem is known. Numerical calculations on the other hand can solve the
TBA equations and therefore extract any measurable quantity.\\
In \cite{zamtba,kl-me} the TBA equations were resolved by an iterative
method. We propose here a multi-grid algorithm, which is considerable faster,
an important fact if many particles are involved.
The heart of the program is the resolution of the coupled integral
equations. Around this core we have designed some utility-programs, in
order to make the tool easier to use. We specialize our application to the
case of diagonal $S$-matrices, see e.g. \cite{zamtba,kl-me,mar-diag}.
As physical quantities we extract the central charge, the dimension of the
perturbing field and the perturbation expansion.
Note though, that one can easily add subroutines calculating other
quantities.
\section{The TBA Equations}
Consider an integrable massive scattering theory on a cylinder.
This implies factorized scattering, and so one
can assume that the wave function of the particles is well described by a
free wave function in the intermediate region of two scattering. Take the
ansatz
$$ \psi ( x_1 \dots x_n ) = e^{ i \sum p_j x_j } \sum _P A(P) \Theta(x_P
) \sa ;$$
$A(P)$ are coefficients of the momenta whose ordering is specified by
$$ \Theta(x_P) = \left \{ \begin{array}{cc}
1 & \mbox{{\rm if}} \sa \sa x_{p_1} < \dots < x_{p_n} \\
0 & \mbox{{\rm otherwise}} \end{array} \right . \sa .
$$
 Let
the permutation $P$ differ
from $P'$ by the exchange of the indices $k$ and $j$. Then
\be
A(P') = S_{kj} (\beta_k-\beta_j) A(P) \sa .
 \label{ex} \ee
We impose antiperiodic boundary conditions for our wave functions,
which provides that two particles cannot have equal momenta, leading to the
condition
\be A(k,p_2, \dots, p_n) = - e^{i p_k L} A(p_2,\dots ,p_n,k)
\sa ,\label{nix} \ee
$L$ being the length of the strip on which we consider the theory.
 comparing (\ref{ex}) and (\ref{nix}), one realizes
\be
e^{i L m_k \sinh \beta_k}
\prod_{j\neq k}S_{kj}(\beta_k-\beta_j) = -1 \sb {\rm for} \sb k=1,2,\dots n
\sa .\ee
We introduce the phase $ \delta_{kj} (\beta _k-\beta_j) \equiv
- i \ln S_{kj}(\beta_k-\beta_j)$. In terms of these the equation become
\be
L m_k \sinh \beta_k + \sum_{j\neq k} \delta_{kj} (\beta_k - \beta_j) = 2
\pi n_k \sb {\rm for} \sa k=1,2,\dots,n \sa , \label{above}
\ee $n_k$ being some integers.
These coupled transcendental equations for the rapidities are called
the Bethe ansatz equations. One tries to solve these equations in the
 thermodynamic limit introducing densities of rapidities for each
particle species and transferring
the equations into integral equations. That is, let $\rho_1^{(a)} (\beta)
= \frac{n} {\Delta \beta}$, where we assume that there are $n$ particles in
the small interval $\Delta \beta$,  be the particle density and
$\rho^{(a)} (\beta)
= \frac{n_k} {\Delta \beta}$ be the level density corresponding to the
particle $a$, then (\ref{above}) becomes
\be
m_a L \cosh \beta + \sum_{b=1}^n\int_{-\infty}^{\infty}
\varphi_{ab}(\beta-\beta')\rho_1^{(a)}
 (\beta') d\beta' =
2 \pi \rho^{(a)}  \sb . \label{dens}
\ee
In order to compute the ground state energy one needs to minimize the free
energy
\be R L f(\rho,\rho_1) =
R H_B (\rho_1) + S (\rho, \rho_1 )\sb ,
\ee
where
$H_B = \sum_a m_a \int \cosh \beta \rho_1^{(a)} d\beta$ and $S$ denotes the
entropy. The extremum condition for a fermionic system\footnote{ We use
the fermionic TBA equations since in diagonal scattering up to now they
turned out to be the relevant ones, see e.g.\cite{zamtba} for the general
theory }
takes the form
\be -r M_a \cosh \beta + \epsilon_a (\beta) =
\sum_{b=1}^n\int_{-\infty}^{\infty} \varphi_{a b}
(\beta - \beta ') \log (1 + e^{-\epsilon_b (\beta)})
\frac{d \beta'}{2 \pi} \sb , \label{tba}
\ee
where we introduced the so-called pseudo-density
$ e^{-\epsilon_a} \equiv \frac{ \rho_1^{(a)} }
{\rho^{(a)} - \rho_1^{(a)} }$, the scaling length
$r=R m_1$ and the rescaled masses $ M_a =\frac{m_a}{m_1}$; $m_1$ is the
lightest particle mass. These coupled integral equations are called the
TBA equations.
 The extremal free energy depends only on the
ratios $\frac{\rho_1^{(a)} }{\rho^{(a)} }$ and is given by
\be f(r) = -\frac r{2 \pi} \sum_{a=1}^n  M_a \int_{-\infty}^{\infty}
\cosh \beta \log(1 + e^{-\epsilon_a (\beta)}
)  d \beta \sb .
\ee

One can extract several physical quantities from the solution of the
TBA-equations (\cite{zamtba,mar-2,kl-me}). Since very little is known about
non-critical systems, one tries to examine the equations in the ultraviolet
limit, which corresponds to $ r \goto 0$, where the underlying field theory
should become a CFT. The central charge is related to the vacuum bulk
energy, and is given by
\be
c(r) = \frac{3 r}{\pi^2} \sum_{a=1}^{n} M_a \int _{-\infty}^{\infty}
 \cosh \beta
\log (1+e^{-\epsilon_a (\beta) }) d \beta \sb .
\ee
Having calculated the central charge one would like to extract the
conformal dimension of the perturbing operator.
For small $r$, one expects that $f(r)$ reproduces the behaviour predicted by
conformal perturbation theory, which in terms of $c(r)$ reads as
\be
c(r)=c -  \frac{3 f_0}{ \pi} r^2 +
\sum_{k=1}^{\infty} f_k r^{yk} \sb ,
\label{exp}
\ee
with possible logarithmic corrections.

The exponent $y$ is related to the perturbing field by $y=2(1-\Delta)$ if
the theory is unitary and by $y=4(1-\Delta)$ if it is non-unitary. The
coefficients are related to correlation functions of the CFT
\cite{zamtba,kl-me}, and even if
one cannot read them off directly, this is an ultimate important check of
the theory.\\
Note that the application chosen is not a limitation of the use of the
program. Also non-diagonal S-matrices (see \cite{zam-nond})
can be treated, since once one has
diagonalized the transfer-matrix also in that case the numerical problem
reduces to solving (\ref{tba}). Further quantities to measure can simply be
added, and also one can study any range of $r$, being a parameter in the
input-data.\\
\section{Description of the Solution Method}
Multi-grid (MG) schemes are known to be the most efficient methods for solving
elliptic boundary value problems. Actually, the underlying idea of treating
the different characteristic length scales of the problem on different grids,
applies successfully also to the numerical
resolution of various other problems, as
the resolution of integral equations \cite{Bie,Hc}. The system of non linear
Fredholm integral equations (\ref{tba}) has been
solved using iterative methods \cite{zamtba,kl-me}. Even if these methods
provide a satisfactory solution in terms of accuracy, the number of iterations
and corresponding computer process (CPU) time required to reach a specified
precision can become excessively large as the number of grid points $N$
increases. Typically a simple one level relaxation
would require $O(N^2 \log N)$ operations. With a multi-grid solution
technique the computing time for integral equations is reduced to $O(N^2)$
\cite{Hc}, and in particular cases to $O(N\log N)$ \cite{Bie},
thus justifying the extra effort in programming.

Now we define our numerical problem and we explain how the multi-grid
scheme works for solving it. In discretising the
TBA equations (\ref{tba}), we use the trapezoidal rule
\cite{Bk} on a grid with mesh size $h$ so that our system yields
\be
\epsilon_a (\beta) =r M_a \cosh \beta +
\frac{h}{2\pi}\sum_{b=1}^{n}\sum_{\beta'\in \Omega_h} w(\beta')
\varphi_{a b} (\beta - \beta ') \log (1 + e^{-\epsilon_b (\beta')})
\sb ,\label{tban}
\ee
$a=1,2,\dots,n $, $\beta\in \Omega_h$,
where $\Omega_h$ is the set of grid points
with grid spacing $h$. The weights are $w(\beta)=1$ unless on the boundary
where $w(\beta)=1/2$.
Now let us introduce a sequence of grids with mesh sizes
$h_1>h_2>...>h_M$, so that $h_{\ell-1}=2 h_{\ell}$. The system (\ref{tban})
with discretisation parameter $h_\ell$ will be denoted as
\be
\epsilon_a^\ell=K_{ab}^\ell(\epsilon_b^\ell)+f_a^\ell \sa ,\sa \sa
a=1,2,\dots,n
\sb ,\label{tbanc}
\ee
where a summation over $b$ is intended and where
\be
K_{ab}^\ell(\epsilon) (\beta) =
\frac{h_\ell}{2\pi}\sum_{\beta'\in \Omega_{h_\ell}} w(\beta')
\varphi_{a b} (\beta - \beta ') \log (1 + e^{-\epsilon (\beta')})\sb .
\ee
Following \cite{Hc} we have applied one (Gauss-Seidel) iteration
to (\ref{tbanc}),
and obtained the approximated solutions $\tilde{\epsilon}_a^\ell , \sa a=1,2,
\dots,n$. We then transfer them onto the next coarser grid,
$\tilde{\epsilon}_a^{\ell-1}=\hat{I}_\ell^{\ell-1}\tilde{\epsilon}_a^\ell$,
where $\hat{I}_{\ell}^{\ell-1}$ is a restriction operator. The coarse grid
equations become
\be
\hat{\epsilon}_a^{\ell-1}=K_{ab}^{\ell-1}(\hat{\epsilon}_b^{\ell-1})
+\hat{f}_a^{\ell-1}
\sa , \sa a=1,2,\dots,n \sb ,
\label{tbancc}
\ee
where
\be
\hat{f}_a^{\ell-1}=I_{\ell}^{\ell-1} f_a^\ell +
\tilde{\epsilon}_a^{\ell-1}-K_{ab}^{\ell-1}(\tilde{\epsilon}_b^{\ell-1})
-I_{\ell}^{\ell-1}(\tilde{\epsilon}_a^\ell-K_{ab}^\ell(\tilde{\epsilon}_b^\ell))
\sb ,\ee
and with $I_{\ell}^{\ell-1}$ another fine-to-coarse grid transfer operator
not necessarily equal to $\hat{I}_{\ell}^{\ell-1}$.
Having obtained the solution of the coarse grid equation
$\hat{\epsilon}_a^{\ell-1}$ the difference
$\hat{\epsilon}_a^{\ell-1}-\tilde{\epsilon}_a^{\ell-1}$ is the coarse-grid
(CG) correction to the fine-grid solution
\be
\tilde{\epsilon}_a^{\ell}\leftarrow \tilde{\epsilon}_a^{\ell}+
\hat{I}_{\ell-1}^{\ell}(\hat{\epsilon}_a^{\ell-1}-\tilde{\epsilon}_a^{\ell-1})
\sb ,\ee
$a=1,2,\dots,n$, and $\hat{I}_{\ell-1}^{\ell}$ is a coarse-to-fine grid
interpolation operator. Finally we perform one relaxation at level $\ell$,
in order to smoothen errors coming from the interpolation procedure.
To solve the system of equations (\ref{tbanc}) we employ a coarse-grid
correction recursively, i.e. equation (\ref{tbancc}) is itself solved by
iteration sweeps combined with a further CG correction.

\section{Numerical Performance}
The algorithm used to solve equations (\ref{tbanc}) is a
non-linear multi-grid (NMGM) method (\cite{Hc}) with full adaptive scheme
(FAS) (\cite{Bg}). The program
can perform different MG variants: V or W cycle and (non linear) nested
iteration, depending on the value of the parameter chosen.
In order to reduce the number of parameters as input we have fixed most of
them, that is the parameter regarding the multi-grid cycle (in order
to optimize it), leaving as an input only the physical ones.
\vskip1.5truecm

\begin{figure}[h]
\begin{picture}(300,240)(10,10)
\special{tban.ps}
\end{picture}
\caption{Evolution of residual error norm with CPU time for a 1-particle
system at $r=0.1$ for different HX: solid line for MG, dashed line
for iteration only.}
\end{figure}

For any scaling length $r$ we use an initial approximation which behaves
like $r M_a \cosh \beta$, wherefrom
the program determines the numerical boundary at
which the kernels vanish and verifies that
the conditions for the existence of (at least) one solution given by the
Schauder's fixed point theorem are satisfied \cite{P}.
Having determined the size of the numerical domain for a given $r$ the number
of levels $M$ is set such that the finest level has a mesh-size $h_M$ of
order of HX, which is one of the input parameters.

\begin{table}[t]
\begin{center}
\begin{tabular}{|c|c|c|}  \hline
{\em no. equations} & \multicolumn{2}{c|}{\em CPU time (secs)} \\
      &   Relax    &  Multi-Grid \\ \hline \hline
  1   &   4        & 3        \\ \hline
  2   &   34       & 22       \\ \hline
  3   &   508      & 331      \\ \hline
  4   &   1230     & 712     \\ \hline
  5   &   2530     & 1320     \\ \hline
\end{tabular}
\end{center}
\caption{A comparison of CPU time required to reach a particular value
of the norm (19), for $r=0.1,\sa {\rm HX}=0.1$.}
\end{table}

We compare the performance of the MG and of a (Gauss-Seidel)
iterative scheme in terms of CPU time in Figure 1, there the different initial
residual error for MG and iterative scheme is due to the set up of the initial
approximated solution in the MG cycle, that is a non-linear nested iteration
which uses a MG cycle itself (see \cite{Hc}). As a norm for the residuals

\be
\tau_a (\beta)=(\epsilon_a-K_{ab}(\epsilon_b)-f_a)(\beta),\sa
\sa \beta \in \Omega_{h_M}\sb,
\ee
we define the norm
\be
\parallel \tau \parallel_M =\max_{1\leq a \leq n} \sqrt{
\sum_{\beta \in \Omega_{h_M}} \tau_a(\beta)^2} \sb.
\ee

\noindent
In order to outline how
the multi-grid algorithm becomes important as the number of particles
increases we give in Table 1 the CPU time required by the two
methods to solve the discretized problem to a value of the residual
norm
\be
\parallel \tau \parallel_M \leq 1 \cdot 10^{-14}\sb .
\label{norm}
\ee

 \section{Structure and Use of the Program}
As already mentioned in the introduction the program consists of two parts:
the core, which resolves the TBA-equations (\ref{tba}) and the periphery,
which on the one hand constructs the kernel, and the initial solution
, and on the other hand extracts from the solution the central charge, the
dimension $y$ and the coefficients $f_i$ of the perturbation expansion
 (\ref{exp}).\\
We specifically designed the program for diagonal scattering theories, that
is we are concerned with scalar $S$ matrices of the form
\be S_{ab} = \prod _i f(\alpha_{ab}^i)\sa,\sb i=1,\dots,
n_{ab}\sa,\sa a,b=1,\dots,n \ee
with $$ f(\alpha)=\frac{\sinh \frac 12 ( \beta + i \pi \alpha)}
{\sinh \frac 12 ( \beta - i \pi \alpha)}$$
$n$ being the number of particles in the theory and $n_{ab}$ is the number
of factors $f_x$ appearing in the $S$-matrix $S_{ab}$
(for a recent review on this subject and many examples, see \cite{sasso}).
The set of the numbers $\alpha$, and  the masses of the theory
are sufficient to resolve the TBA-equations. \\
As input-data we have therefore three data files: TBA.DAT containing
general information about the range of $r$, the mesh-size
and the number of particles,
and the file ALPHA.DAT containing the values of $\alpha$ and MASS.DAT
containing the values of the mass of the particles.
{}From the input the program then constructs the kernel, and the
initial approximation, which are then used by the multi-grid algorithm.
The routine returns the solutions encoded in a matrix array Q which is stored
in the output file SOL.DAT. The physical parameters are then calculated
in the appropriate following subroutines, see the flow-diagram.

We discuss here specific structure of the algorithm in terms of
a simple example. Consider the $S$-matrix

\be
S_{11}=f_{\frac{2}{5}} f_{\frac{3}{5}},\sa
S_{12}=S_{21}=f_{\frac{1}{5}} f_{\frac{2}{5}} f_{\frac{3}{5}} f_{\frac{4}{5}},
\sa S_{22}=f_{\frac{1}{5}} f_{\frac{4}{5}} (f_{\frac{2}{5}} f_{\frac{3}{5}})^2,
\ee
being known to describe the minimal model $\M_{2,7}$, (i.e. $c_{eff}=
\frac{4}{7}$) perturbed by the field with dimension $\Delta=-\frac{3}{7}$.
The masses of the two particles are
\be
M_1=1 \sa, \sa \sa M_2= 2 \cos (\frac{\pi}{5}).
\ee
Then the input-file ALPHA.DAT should read as

{\tt
\begin{tabular}{rr}
 2.0 &  5.0 \\
 3.0 &  5.0 \\
-2.0 &  1.0 \\
 1.0 &  5.0 \\
 4.0 &  5.0 \\
 2.0 &  5.0 \\
 3.0 &  5.0 \\
-2.0 &  1.0 \\
 2.0 &  5.0 \\
 3.0 &  5.0 \\
 2.0 &  5.0 \\
 3.0 &  5.0 \\
 1.0 &  5.0 \\
 4.0 &  5.0 \\
-2.0 &  1.0 \\
\end{tabular}
}

\noindent
Every factor $f_{\alpha}$ is characterized by two numbers which are the
nominator and denominator of $\alpha$. As a field separator between values
$\alpha$ belonging to different $S$-matrix elements one gives a number
smaller than $-1$. Since the $S$-matrix is symmetric the program reads only
the elements $S_{1,1},S_{1,2},\dots,S_{1,n};S_{2,2},\dots,S_{2,n};\dots;
S_{n-1,n-1},S_{n-1,n};S_{n,n}$.

The file MASS.DAT contains two values:

{\tt
\begin{tabular}{l}
  1.0                 \\
  1.618033988749895   \\
\end{tabular}
}

\noindent
Finally the file TBA.DAT contains in our example the following data

{\tt \begin{tabular}{l}
15,2 \\
1.0d-14,1.0d-2\\
0,1\\
30,0.01,0.01\\
20.0,7.0,0,0,4\\
4.,7.\end{tabular}
} corresponding to {\tt
\begin{tabular}{l}
I1,I2\\
ZERO,HX\\
NREL,IWRITE\\
MAX,STEP,R0\\
YN,YD,NY,NCEX,MFIT\\
CEXN,CEXD\end{tabular} }

\noindent
These parameters are:
\begin{itemize}
\item[I1,I2:]
 corresponding lengths of the files ALPHA.DAT and MASS.DAT (I2 coincides
with the number of equations of our system, NSYS in the program);
\item[ZERO:] the order of the value of the residual norm to be reached;
HX the finest grid size;
\item[NREL:]
if NREL$=0$ multi-grid algorithm, if NREL$=1$ only relaxation is done
\item[MAX:]
 the number of different radii $r$ to be used; R0 is the smallest radius, and
STEP give the step-size of the radius in the cycle;
\item[YN, YD:]
are the numerator and the denominator of
the exact exponent if available, otherwise dummy numbers; if NY$=0$
the exact exponent is used in the fitting procedure, if NY$=1$ the estimated
$y$ is used; if MFIT$> 0$ the program calculates the first MFIT coefficients
$f_i$ (in this case must be MAX $\geq{\rm  MFIT}+5$). Finally
if NCEX$=0$, the exact charge (cexact=CEXN/CEXD)
is used for the fitting procedure.
\end{itemize}

Some comments: The CPU time is mainly determined
from the parameters NSYS, being the dimension of the system, MAX and HX.
MAX is chosen corresponding
to the physical parameters one wants to compute: in order to get a sensible
result for the dimension $y$ it is enough to use MAX $\sim 5$, whereas if one
wants to calculate the coefficients $f_i$ a larger number is required; for
example with MAX$\geq 10$ one can get $f_1$ up to $O(10^{-5})$; the more
data is used the better the fit-procedure \cite{NR} works and more
coefficients can be obtained. HX determines the error of the integration
routines calculating $c(r)$. In any case HX=$0.01$ should be sufficient.

The program produces four files of output. If IWRITE=1 only the physical
informations, that is $c$, $y$ and $f_i$ are stored in the file OUTPUT.DAT
(see test-run output). When IWRITE=0 the program generates, in addition, the
file RES.DAT which containes technical data and the evolution of the value of
the residuals during the iterations. Together with RES.DAT other two files are
written: TIME.DAT which containes the logarithm of the residual norm and the
actual CPU time used, and SOL.DAT (where the solutions for any $r$ are stored).

\section{Test-Run Output}
The following file was produced using the input files described in the
last section:
\begin{enumerate}
\item OUTPUT.DAT
\begin{verbatim}
   computed cexact =    .57142857D+00
   r=   .100000000D-01  central charge=   .571403656E+00
   r=   .200000000D-01  central charge=   .571329513E+00
   r=   .300000000D-01  central charge=   .571206952E+00
   r=   .400000000D-01  central charge=   .571036718E+00
   r=   .500000000D-01  central charge=   .570819520E+00
   r=   .600000000D-01  central charge=   .570556042E+00
   r=   .700000000D-01  central charge=   .570246948E+00
   r=   .800000000D-01  central charge=   .569892886E+00
   r=   .900000000D-01  central charge=   .569494492E+00
   r=   .100000000D+00  central charge=   .569052390E+00
   r=   .110000000D+00  central charge=   .568567192E+00
   r=   .120000000D+00  central charge=   .568039505E+00
   r=   .130000000D+00  central charge=   .567469925E+00
   r=   .140000000D+00  central charge=   .566859042E+00
   r=   .150000000D+00  central charge=   .566207440E+00
   r=   .160000000D+00  central charge=   .565515695E+00
   r=   .170000000D+00  central charge=   .564784379E+00
   r=   .180000000D+00  central charge=   .564014059E+00
   r=   .190000000D+00  central charge=   .563205295E+00
   r=   .200000000D+00  central charge=   .562358645E+00
   r=   .210000000D+00  central charge=   .561474660E+00
   r=   .220000000D+00  central charge=   .560553889E+00
   r=   .230000000D+00  central charge=   .559596875E+00
   r=   .240000000D+00  central charge=   .558604159E+00
   r=   .250000000D+00  central charge=   .557576277E+00
   r=   .260000000D+00  central charge=   .556513761E+00
   r=   .270000000D+00  central charge=   .555417143E+00
   r=   .280000000D+00  central charge=   .554286946E+00
   r=   .290000000D+00  central charge=   .553123695E+00
   r=   .300000000D+00  central charge=   .551927909E+00
  error in extrapolation  .3365E-09
  estimated exponent       .285714287D+01
  theoretical exponent     .285714286D+01
  estimated dimension of the corresponding operator
  for a     unitary theory: DELTA=    .285714D+00
  for a non-unitary theory: DELTA=   -.428571D+00
   fitted f_i
  f( 1)= .9643967331341316D-01
  f( 2)=-.1538311769518447D-02
  f( 3)= .6222166295705919D-04
  f( 4)=-.3197939259001748D-05
   chi-square value of the fitting=  .5164E-29
   total cpu time (secs)   .113E+06
\end{verbatim}
\end{enumerate}
\noindent
\section{Conclusions}

We presented a multi-grid scheme for the resolution of the thermodynamic
Bethe ansatz equations. The TBA is a means to describe the finite
temperature effects of relativistic factorized scattering theories.
Our program is specifically designed for theories having a scalar
$S$-matrix. These theories exhibit a unique form, and the only input needed
in order to carry out the TBA are the locations of the poles and zeros of
the single $S$-matrix elements.

The program calculates the central charge and in the ultra-violet limit
the dimension of the perturbing field and the coefficients of the
perturbation expansion.
These are the most crucial tests in verifying a conjectured $S$-matrix.
It should not be difficult for the user to add subroutines calculating
other physical quantities, as for example for magnetic systems the moments
of the total magnetization, or the convergence-region of the perturbation
series in (\ref{exp}) \cite{zamtba,kl-me}.

In order to get sensible results for the physical quantities
one needs to resolve the integral equations with the highest possible
accuracy. This unfortunately renders the calculation extremely time consuming.
Therefore the use of an efficient Multi-Grid algorithm gives the
possibility to reach high accuracy in the computation together with a sensible
reduction of the CPU time, in confrontation with standard
iterative techniques.

{\bf Acknowledgments}\\
We are grateful to G. Mussardo and M. Martins for encouragement and
discussions, and to A. Lanza for introducing us to the multi-grid
method.


\newpage
\begin{thebibliography}{99}
\bibitem{zam} A.B. Zamolodchikov, {\em JETP Letters} {\bf 46} (1987) 160;
{\em Int. Journ. Mod. Phys.} {\bf A3} (1988) 743;
{\em Advanced Studies in Pure Mathematics} {\bf 19} (1989) 641.
\bibitem{zamtba} Al.B. Zamolodchikov, {\em Nucl. Phys.} {\bf B342} (1990) 695.
\bibitem{kl-me} T. Klassen, E. Melzer {\em Nucl. Phys.} {\bf B350} (1990) 635.
\bibitem{mar-2} M.J. Martins, {\em Phys. Rev. Lett.} {\bf 67} (1991) 419.
\bibitem{mar-diag} M.J. Martins, {\em Phys. Lett.} {\bf B240} (1990) 404;
P. Christe, M.J. Martins, {\em Mod. Phys. Lett.} {\bf A5} (1990) 2189;
A. Koubek, M.J. Martins, G. Mussardo {\em Nucl. Phys. } {\bf B368}
(1992) 591.
\bibitem{zam-nond} Al.B. Zamolodchikov, {\em Nucl. Phys.} {\bf B358}
(1991) 497.
\bibitem{sasso} G. Mussardo, {``Off-critical statistical models: factorized
scattering theories and Bootstrap program''}, {\em Phys. Rep.}, to appear.
\bibitem{Bie} A. Brandt, A.A. Lubrecht,
 {\em J. Comp. Phys.} {\bf 90} (1990) 348.
\bibitem{Bg} A. Brandt, {\em Multi-grid techniques: 1984 guide with
applications to fluid dynamics} (GMD-Studien. no 85,
St. Augustin, Germany, 1984).
\bibitem{Bk} C.T.H. Baker,{\em The numerical treatment of integral equations}
(Oxford University Press, London, 1977) p.110.
\bibitem{Hc} W. Hackbusch,{\em Multi-Grid Methods and Applications}
(Springer-Verlag Berlin, Heidelberg, 1985) p.305.
\bibitem{P} W. Pogorzelski,{\em Integral Equations and their Applications,
Vol. I} (PWN-Polish Scientific Publishers, Warsaw, 1966) p.201.
\bibitem{NR} W.T. Vetterling, S.A. Teukolsky, W.H. Press, B.P. Flannery,
{\em Numerical Recipes: The art of scientific computing} (Cambridge University
Press, New York, 1985) p.521.
\end{thebibliography}
\end{document}